\documentclass[12 pt]{article} 
\usepackage{times}
\usepackage{authblk,float}
\usepackage{graphicx}
\usepackage{bm}
\usepackage[pagebackref=false,colorlinks,linkcolor=blue,citecolor=blue]{hyperref}
\usepackage[margin=1.1 in]{geometry}
\usepackage{booktabs} 
\usepackage{array} 
\usepackage{paralist} 
\usepackage{verbatim} 
\usepackage{subfig} 
\usepackage{fancyhdr} 
\usepackage{sectsty}
\usepackage[nottoc,notlof,notlot]{tocbibind} 
\usepackage[titles,subfigure]{tocloft} 
\usepackage{cite}
\usepackage{amsmath}
\usepackage{amsfonts}
\usepackage{amssymb}
\usepackage[usenames, dvipsnames]{color}
\usepackage{float}

\begin{document}

\title{ \textmd{The Dirac oscillator in a spinning  cosmic string spacetime}}

\author{\large \textit {Mansoureh Hosseinpour}$^{\ 1}$\footnote{E-mail:hosseinpour.mansoureh@gmail.com} ,~\textit{Hassan Hassanabadi}$^{\ 1}$\footnote{E-mail:hha1349@gmail.com}, \\~\textit {   {Marc} de Montigny}$^{\ 2}$\footnote{E-mail:mdemonti@ualberta.ca}
	\\
	
	\small \textit {$^{\ 1}$ Faculty of Physics, Shahrood University of Technology, Shahrood, Iran}\\
	\small \textit{P.O. Box 3619995161-316}\\\small \textit {$^{\ 2}$ {Facult\'e} Saint-Jean, University of Alberta, Edmonton, Alberta, T6C 4G9, Canada}}

\date{}

\maketitle

\begin{abstract}
  \textmd{ We examine the effects of gravitational fields produced by topological defects on a Dirac field and a Dirac oscillator in a spinning cosmic string spacetime.  We obtain the eigenfunctions and the energy levels of the relativistic field in that background and consider the effect of various parameters, such as the frequency of the rotating frame, the oscillator's frequency, the string density and other quantum numbers.  }
\end{abstract}

\begin{small}
\textmd{\textbf{Key Words}: Dirac oscillator, cosmic strings }
\end{small}
\\\\
\begin{small}
\textmd{\textbf{PACS}: 03.65.Pm, 04.62. +v, 11.27.+d  }

\end{small}

\section{Introduction}

The Dirac equation with interactions linear in the coordinates was initially studied in Refs. \cite{1,2,3}. Such a system was referred to as a `Dirac oscillator' in Ref. \cite{4}, because, in the non-relativistic limit, it behaves as a harmonic oscillator with a strong spin-orbit coupling term.  This model describes the dynamics of a harmonic oscillator for spin-half particles and is obtained by introducing a non-minimal prescription into the free Dirac equation \cite{4}. It was observed that the Dirac oscillator interaction is a physical system which can be interpreted as the interaction of the anomalous magnetic moment with a linear electric field \cite{5,6}. The authors of Ref. \cite{5} also established the conformal invariance of the Dirac oscillator  and the authors of Ref. \cite{6} examined its covariance properties, applied the Foldy-Wouthuysen and the Cini-Toushek transformations. As a relativistic quantum mechanical system, the Dirac oscillator has been widely studied and, because it is an exactly solvable model, several investigations have been performed in its theoretical framework. 

Although the Dirac oscillator is normally utilized within the context of many-body theory, relativistic quantum mechanics and quantum chromodynamics (such as the interaction between quarks as well as the confining part of the phenomenological Cornell potential), the Dirac oscillator and related models have been applied in many other contexts as well, such as quantum optics \cite{7,8,9}, supersymmetry \cite{5,11,12}, nuclear reactions \cite{13}, the hadronic spectrum (with the two-body Dirac oscillator) \cite{14,15}, the Clifford algebra \cite{16,17}, non-commutative space\cite{18,19}, thermodynamic properties \cite{20}, Lie algebras \cite{21}, supersymmetric (non-relativistic) quantum mechanics \cite{22}, the supersymmetric path-integral formalism \cite{23}, chiral phase transitions in presence of a constant magnetic field { \cite{8},} the relativistic Landau levels in presence of external magnetic field \cite{25}, the Aharonov-Bohm effect\cite{26}, and  condensed matter physics phenomena and graphene \cite{27}. Similar studies for the Duffin-Kemmer-Petiau (DKP) oscillator, which is analogous to the Dirac oscillator for spinless and spin-one particles, are in Refs. \cite{28,29}. Finally, let us mention many studies of the Dirac oscillator with topological defects and cosmic string spacetimes in Refs. \cite{DO1,DO2,DO3,DO4,DO5,DO6,DO7} and analogous investigations for scalar fields in Refs. \cite{SCA1,SCA2,SCA3,SCA4,SCA5}. Some studies of relativistic oscillators are in Refs. \cite{RO1,RO2,RO3,RO4,RO5,RO6,RO7,RO8}.

In this work,  we examine the relativistic quantum dynamics of Dirac oscillator on the curved spacetime of a rotating cosmic string. From the corresponding Dirac equation, we analyze the influence of the topological defect on the equation of motion, the energy spectrum and the wave-functions. An analogous study for the Klein-Gordon equation is in Ref. \cite{Cunha2016}. In Sec. 2, we write down the covariant Dirac equation without oscillator in a spinning cosmic string spacetime, and find its wave-functions and energy eigenvalues. In Sec. 3, we present the covariant Dirac oscillator in  the same spacetime and obtain the wave-functions and energy spectrum. We present concluding remarks in Sec. 4.


\section{Dirac equation in the cosmic string spacetime  }

The spacetime generated by a spinning cosmic string without internal structure, or `ideal' spinning cosmic string, can be obtained from the line element 
\begin{align}\label{1}
ds^2=-dT^2+dX^2+dY^2+dZ^2
\end{align}
by applying the coordinate transformation
\begin{align}\label{2}
\begin{gathered}
  T = t + a{\alpha ^{ - 1}}\varphi,  \hfill \\
  X = r{\cos}\varphi,  \hfill \\
  Y = r{\sin}\varphi,  \hfill \\
  \varphi  = \alpha \varphi', \hfill \\ 
\end{gathered} 
\end{align}
which leads to (see also Refs. \cite{30,31,32,33})
\begin{eqnarray}\label{3}
d{s^2} &=&  - {(dt + ad\varphi )^2} + d{r^2} + {\alpha ^2}{r^2}d{\varphi ^2} + d{z^2}\nonumber\\	
&=&  - d{t^2} - 2adtd\varphi\   + d{r^2} + ({\alpha ^2}{r^2} - {a^2})d{\varphi ^2} + d{z^2},
\end{eqnarray}
where $- \infty  < z < \infty$ , $r \ge 0$ and $0 \le \varphi  \le 2\pi$.  We work with units such that $c=1$. The angular parameter $ \alpha$ runs in the interval $ \left( {0,1} \right]$ and  is related to the linear mass density $\mu$ of the string by $\alpha  = 1 - 4\mu$, and to the deficit angle by $ \gamma=2\pi(1-\alpha) $. We have also $a = 4Gj$ where $G $ is the universal gravitation constant and $j$ is the angular momentum of the spinning string; thus $a$ is a length that represents the rotation of the cosmic string. Note that in this case, the source of the gravitational field of a spinning cosmic string possesses angular momentum and the metric \eqref{3} has an off-diagonal term involving time and space.

The Dirac equation for a field $\Psi$ of mass $M$ in the cosmic string spacetime {described by Eq. \eqref{3} reads\cite{34,35,36,37}
\begin{align}\label{4}
({\rm{i}}{\gamma ^\mu }\left( x \right){\nabla _\mu } - {M})\Psi \left( x \right) = 0,
\end{align}
with the covariant derivative
\begin{align}\label{5}
{\nabla _\mu } = {\partial _\mu } + {\Gamma _\mu }\left( x \right),
\end{align}
and the spinorial affine connections, 
\begin{align}\label{6}
{\Gamma _\mu } = \frac{1}{2}{\omega _{\mu \bar a\bar b}}\left[ {{\gamma ^{\bar a}},{\gamma ^{\bar b}}} \right],
\end{align}
where ${{\gamma ^{\bar a}}}$ denotes the standard Dirac matrices in Minkowski spacetime  with metric $\eta_{\bar a \bar b}=\left(-1,+1,+1,+1\right)$, and $\omega _{\mu \bar a \bar b}$ is the spin connection, given by
\begin{align}\label{7}
{\omega _{\mu \bar a \bar b}} = {\eta _{ \bar a \bar c}}e_{\; \nu}^{\bar c}e_{\bar b}^\sigma \Gamma _{\sigma \mu }^\nu  - {\eta _{\bar a \bar c}}e_{ \bar b}^\nu e_\nu ^{ \bar c}.
\end{align}
We use greek indices $\mu$, $\nu$, etc. for the curved spacetime, and bar latin indices $\bar a$, $\bar b$, etc. in Minkowski spacetime.

As discussed in Refs. \cite{CaiPapini,Vilenkin1985}, the spin connection allows us to construct a local frame through the tetrad basis which gives the spinors in the curved spacetime. The Christoffel symbols of the second kind, $\Gamma _{\mu \nu }^\sigma $, can be obtained from
\begin{align}\label{8}
 \Gamma _{ij}^\mu  = \frac{1}{2}{g^{\mu k}}{\left(\frac{{\partial {g_{ik}}}}{{\partial {q_j}}} + \frac{{\partial {g_{jk}}}}{{\partial {q_i}}} - \frac{{\partial {g_{ij}}}}{{\partial {q_k}}}\right)},
\end{align}
with coordinates $(q_1,q_2,q_3)$.  With the metric in Eq. (\ref{3}), the non-null components of Christoffel symbols are  
\begin{align}\label{9}
\Gamma _{r\varphi }^t = \Gamma _{\varphi r}^t = \frac{{ - a}}{r},\,\,\,\,\,\,\,\,\Gamma _{\varphi \varphi }^r =  - r{\alpha ^2},\,\,\,\,\,\,\,\Gamma _{r\varphi }^\varphi  = \Gamma _{\varphi r}^\varphi  = \frac{1}{r}.
\end{align}

We can build the local reference frame through a non-coordinate basis with components $e_\mu ^{{\bar a}}$ called tetrads or vierbeins which form our local reference frame. With the line element \eqref{3}, we can use tetrads $e_{{\bar a}}^\mu $  and $e_\mu ^{{\bar a}}$ (obtained in Ref. \cite{32}) as follows
\begin{align}\label{10}
e_\mu ^{{\bar a}} = \left( {\begin{array}{*{20}{c}}
  1&0&a&0 \\ 
  0&{\cos\varphi}&{ - r\alpha {\sin\varphi}}&0 \\ 
  0&{\sin\varphi}&{ - r\alpha {\cos\varphi} }&0 \\ 
  0&0&0&1 
\end{array}} \right),\quad e_{{{\bar a}}}^\mu  = \left( {\begin{array}{*{20}{c}}
  1&{\frac{a{\sin\varphi}}{{r\alpha }}}&{\frac{-a{\cos\varphi}}{{r\alpha }}}&0 \\ 
  0&{\cos\varphi}&{\sin\varphi}&0 \\ 
  0&{\frac{-{\sin\varphi}}{{r\alpha }}}&{\frac{\cos \varphi }{{r\alpha }}}&0 \\ 
  0&0&0&1 
\end{array}} \right).
\end{align}
The vierbeins satisfy the orthonormality conditions
\begin{align}\label{11}
\begin{array}{l}
e_{ {{\bar a}}}^\mu \left( x \right)e_\nu ^{ {{\bar a}}}\left( x \right) = \delta _\nu ^\mu, \\
e_\mu ^{ {{\bar a}}}\left( x \right)e_{ {{\bar b}}}^\mu \left( x \right) = \delta { _{\bar b}^{\bar a}},
\end{array}
\end{align}
and
\begin{align}\label{12}
{g_{\mu \nu }}\left( x \right) = e_\mu ^{  {\bar a} }\left( x \right)e_\nu ^{  {\bar b} }\left( x \right){\eta _{^{{\bar a}{\bar b}}}}.
\end{align}\\
The non-null components of the spin connection are
\begin{align}\label{13}
\omega _\varphi ^{ { {\bar1}{\bar2}}} =  - \omega _\varphi ^{{ {\bar2}{\bar1}}} =1- \alpha, 
\end{align}
and the only non-vanishing spinorial affine connection is 
\begin{align}\label{14}
{\Gamma _\varphi } = -\frac{{ 1}}{2}(1 - \alpha ){\gamma ^{{ \bar1}}}{\gamma ^{{ \bar2}}}.
\end{align}

The generalized Dirac matrices $\gamma^\mu$ in curved spacetime are related to their Minkowski counterparts via
\begin{align}\label{15}
{\gamma ^\mu }\left( x \right) = e_{ {\bar a}}^\mu \,{\gamma ^{ {\bar a}}}.
\end{align}
In terms of the Minkowski flat spacetime coordinates, these matrices can be cast into the form
\begin{align}\label{16}
{\gamma ^ {{ {\bar 0}}} } = \left( {\begin{array}{*{20}{c}}
1&0\\
0&{ - 1}
\end{array}} \right),{{\gamma ^{ {\bar i}}} = \left( {\begin{array}{*{20}{c}}
0&{{\sigma ^i}}\\
{-{\sigma ^i}}&0
\end{array}} \right)},\qquad { i=1, 2, 3,}
\end{align}
where ${{\sigma ^i}}$ are the Pauli matrices:
\begin{equation}\label{Pauli}
\sigma^1= \left( \begin{array}{cc}
0&1\\
1&0 \end{array}\right),\quad
\sigma^2= \left( \begin{array}{cc}
0&-i\\
i&0 \end{array}\right),\quad
\sigma^3= \left( \begin{array}{cc}
1&0\\
0&-1 \end{array}\right).\quad
\end{equation}

The matrices ${\gamma ^\mu }\left( x \right)$ in Eq. \eqref{6} are (see Ref. \cite{32})
\begin{align}\label{17}
\begin{gathered}
  {\gamma ^r} = {\gamma ^1} = e_{\bar a}^1{\gamma ^{\bar a}} = {\gamma ^{\bar 1}}\cos \varphi  + {\gamma ^{\bar 2}}\sin \varphi , \hfill \\
   {\gamma ^z} = e_{\bar a}^z{\gamma ^{\bar a}} = {\gamma ^3}, \hfill \\ 
  {\gamma ^2} = e_{\bar a}^2{\gamma ^{\bar a}} = \frac{{{\gamma ^\varphi }}}{{r\alpha }}, \hfill \\
  {\gamma ^\varphi } = e_{\bar 0}^\varphi {\gamma ^{\bar 0}} + e_{\bar 1}^\varphi {\gamma ^{\bar 1}} + e_{\bar 2}^\varphi {\gamma ^{\bar 2}} =  - {\gamma ^{\bar 1}}\sin \varphi  + {\gamma ^{\bar 2}}\cos \varphi , \hfill \\
     {\gamma ^t}= {\gamma ^0} = e_{\bar a}^0{\gamma ^{\bar a}} =  {\gamma ^{\bar 0}} + \frac{{a\sin \varphi }}{{r\alpha }}{\gamma ^{\bar 1}} - \frac{{a\cos \varphi }}{{r\alpha }}{\gamma ^{\bar 2}} = {\gamma ^{\bar 0}} - \frac{a}{{r\alpha }}{\gamma ^\varphi }. \hfill \\
\end{gathered} 
\end{align}

The $z$-translation symmetry of Eq. (\ref{3}) allows us to reduce the four-component Dirac equation \eqref{2} to two two-component spinor equations. 
 We shall also take $p_z = 0$ and $z = 0$.  We find that Eq. \eqref{14} yields
\begin{align}\label{19}
\Gamma_\varphi=i \frac{(1-\alpha)}{2}s\sigma^{z}.
\end{align}

If we substitute the curved-spacetime gamma matrices of Eq. (\ref{17}) into Eq. (\ref{4}), we find 
\begin{equation}\label{A}
\left(i\gamma^t\partial_t+i{\vec\gamma}\cdot{\nabla}+i\gamma^\mu\Gamma_\mu\right)\Psi=\left[i\left({\gamma^{\bar 0}} - \frac{a}{{r\alpha }}{\gamma^\varphi}\right)\partial_t+i{\vec\gamma}\cdot{\nabla}+i\gamma^\mu\Gamma_\mu\right]\Psi=M\Psi
\end{equation}
Since we work in the plane $z=0$, we effectively have ${\vec\gamma}=\left(\gamma^r,\gamma^\varphi\right)$ and $\nabla\rightarrow\nabla_\alpha\equiv{\hat r}\frac{\partial}{\partial r}+\frac{\hat\varphi}{\alpha r}\frac{\partial}{\partial\varphi}$.  Since Eq. (\ref{14}) is the only non-zero component of $\Gamma_\mu$, we write $\gamma^\mu\Gamma_\mu={\vec\gamma}\cdot{\vec\Gamma}$, with ${\vec\Gamma}=(\Gamma_r,\Gamma_\varphi)=(0,\Gamma_\varphi)$.
Next we multiply Eq. (\ref{A}) with $\gamma^{\bar 0}$,
\begin{equation}\label{AA}
\left[i\left(\gamma^{\bar 0}{\gamma^{\bar 0}} - \frac{a}{{r\alpha }}\gamma^{\bar 0}\gamma^{\varphi}\right)\partial_t-\gamma^{\bar 0}M\right]\Psi=\gamma^{\bar 0}{\vec\gamma}\cdot\left[-i{\nabla_\alpha}-i{\vec\Gamma}\right]\Psi.
\end{equation}
From Eq. (\ref{16}), we see that ${\gamma^{\bar 0}}{\gamma^{\bar 0}}=1$ and we define $\alpha^i\equiv{\gamma^{\bar 0}}{\gamma^i}=
\left(\begin{array}{cc}
0&{{\sigma ^i}}\\
{{\sigma ^i}}&0\end{array}\right)$.
Since the interaction is time-independent, one can write 
\begin{equation}\label{n22}
\Psi(t,r,\varphi) = {e^{ - i(Et - m\varphi )}}\Psi (r), 
\end{equation}
where $E$ is the energy of the fermion, so that $i\partial_t\Psi=E\Psi$, and $\partial_\varphi\Psi=im\Psi$, where the quantum number $m$ is related to the $z$-component of the total angular momentum operator, with half-integer values (see e.g. Eq. (15) in Ref. \cite{DO1} and Ref. \cite{32}).  Then Eq. (\ref{AA}) becomes
\begin{equation}\label{AAA}
\left(E -\gamma^{\bar 0}M\right)\Psi=\gamma^{\bar 0}{\vec\gamma}\cdot\left(-i{\nabla_\alpha}-i{\vec\Gamma}\right)\Psi+ \frac{aE}{{r\alpha }}\alpha^{\bar 2}\Psi.
\end{equation}

We write the radial wave-function as
\begin{align}\label{20}
\Psi (r) = \left( {\begin{array}{*{20}{c}}
{\phi(r)}\\
{{\chi}(r)}
\end{array}} \right),
\end{align}
where $\phi(r)$ and $\chi(r)$ are two-spinors. 
If we write 
\[
\gamma^{\bar 0}{\vec\gamma}\cdot{\hat\varphi}=\alpha^\varphi,
\]
so that $ \frac{aE}{{r\alpha }}\alpha^{\bar 2}=\gamma^{\bar 0}{\vec\gamma}\cdot\left(\frac{aE}{{r\alpha }}{\hat\varphi}\right)$,  }
and define the generalized momentum $\vec \pi$ as follows,
\begin{align}\label{30}
\vec{\pi}=-i\nabla_\alpha-i{\vec\Gamma}+\frac{aE}{\alpha r}\hat\varphi,
\end{align}
then Eq. (\ref{AAA}) leads to the coupled equations 
\begin{align}\label{30new}
\begin{gathered}
  \left( {E - M} \right)\phi (r) = {\vec\sigma}\cdot{\vec\pi} \ \chi (r), \hfill \\
  \left( {E + M} \right)\chi (r) ={\vec\sigma}\cdot{\vec\pi} \ \phi (r). \hfill \\ 
\end{gathered} 
\end{align}
We substitute the second equation of Eq. (\ref{30new}) into the first and obtain
\begin{equation}
\left(E^2-M^2\right)\phi=\left({\vec\sigma}\cdot{\vec\pi}\right) \left({\vec\sigma}\cdot{\vec\pi}\right) \phi={\vec\pi}\cdot{\vec\pi}\phi.
\label{n25}\end{equation}
With $\Gamma_\varphi$ from Eq. (\ref{19}), we find that the explicit form of Eq. (\ref{30}) is
\[
\vec{\pi}=-i\nabla_\alpha+\left(\frac1{2\alpha r}(1-\alpha)s+\frac{aE}{\alpha r}\right)\hat\varphi =: -i\nabla_\alpha+\frac\xi r\hat\varphi,
\]
where
\begin{equation}
\xi  \equiv  \frac1{2\alpha}(1-\alpha)s+\frac{aE}\alpha.
\label{xi}\end{equation}
We find
\begin{equation}\label{pidotpi}
{\vec\pi}\cdot{\vec\pi}\phi=\left(-i\nabla_\alpha+\frac\xi r\hat\varphi\right)\left(-i\nabla_\alpha+\frac\xi r\hat\varphi\right) \phi=
\left(-\nabla_\alpha^2+\frac{\xi^2}{r^2}-2i\frac\xi r\frac 1{\alpha r} \partial_\varphi\right)\phi.
\end{equation}
When we substitute this into Eq. (\ref{n25}), use Eq. (\ref{n22}) and 
\(
\nabla_\alpha^2=\frac{\partial^2}{\partial r^2}+\frac 1r\frac \partial{\partial r}+\frac 1{\alpha^2 r^2}\frac{\partial^2}{\partial \varphi^2},
\)
we obtain
\begin{equation}\label{AF}
\frac{d^2\phi}{d r^2}+\frac 1r\frac {d\phi}{d r}+\left[E^2-M^2-\left(\frac{m^2}{\alpha^2 r^2}+\frac{\xi^2}{r^2}+2m\frac\xi {\alpha r^2}\right)\right]\phi=0
\end{equation}

Note that $\phi(r)$ (and $\chi(r)$) is an eigenfunction of $\sigma^3$ with eigenvalues $\pm 1$, so we can write $\phi_s=\left(\phi_+, \phi_-\right)^T$ with $\sigma^3\phi_s=s\phi_s$, $s=\pm 1$.  Therefore the action of $\xi$ on $\phi_s$ is the reason why we replaced $\sigma^3$ by $s$ in Eq. (\ref{xi}). Since, in Eq. (\ref{AF}), we have $\left(\frac{m^2}{\alpha^2 r^2}+\frac{\xi^2}{r^2}+2m\frac\xi {\alpha r^2}\right)=\frac 1{r^2}\left(\frac m\alpha+\xi\right)^2$, we can express Eq. (\ref{AF}) as 
\begin{align}\label{21}
\left[\frac{{{d^2}}}{{d{r^2}}} + \frac{1}{r}\frac{d}{{dr}} + {\kappa ^2} - \frac{{{\eta ^2}}}{{{r^2}}}\right]\phi \left( r \right) = 0,
\end{align}  
where 
\begin{align}\label{22}
\kappa^2={E^2} - {M^2},\qquad 
\eta ^2 =\left(\frac m\alpha+\xi\right)^2.
\end{align}

By performing the change of variable $r =\frac x\kappa$,  we can write Eq. \eqref{21} in the form 
\begin{align}\label{25}
{\phi} ^{\prime \prime }(x) +{\frac1x{{ {\phi }'(x)}}  } + \left( {1 - \frac{{{\eta^2}}}{{{x^2}}}} \right){ {\phi (x)}} = 0.
\end{align}
The physical solution of Eq. \eqref{25} is given by the Bessel function of the first kind, $J_\eta(x)$ so that the radial wave-function is given by 
\begin{align}\label{27}
\phi _\eta\left( r \right) = {J_\eta }(\kappa r).
\end{align}
We will not discuss these solutions any further, because we will discuss in detail the more general Dirac oscillator in the next section.


\section{ The Dirac oscillator in cosmic string background}

In this section, we  turn on the Dirac oscillator interaction by adding the non-minimal substitution term,
\begin{align}\label{28}
{\partial _r} \to {\partial _r} + M\omega\vec r \beta{,}
\end{align}
where  $\beta=\gamma^0$, $\vec r=\hat r r$ is the position vector, and $\omega$ is the oscillator's frequency. (In the presence of a constant magnetic field $B_0$, $\omega=\frac{eB_0}{2M}$ is the so-called cyclotron frequency of the oscillator.) We substitute Eq. (\ref{28}) into  Eq. \eqref{4} and obtain
\begin{align}\label{29}
\left[ {{\vec\alpha} \cdot\left( {\vec \pi  - iM\omega  \gamma^0 \hat rr} \right)} \right]\Psi (r) = \left( {E - {\gamma^{\bar 0}}M} \right)\Psi(r),
\end{align} 
where  ${\vec \pi}$ is given in Eq. \eqref{30} and  $\Psi(r)$ in Eq. \eqref{20}. We obtain the counterpart of Eq. \eqref{30new}:
\begin{align}\label{30newbis}
\begin{gathered}
  \left( {E - M} \right)\phi (r) = {\vec\sigma}\cdot\left({\vec\pi}+iM\omega{\hat r}r\right) \ \chi (r), \hfill \\
  \left( {E + M} \right)\chi (r) ={\vec\sigma}\cdot\left({\vec\pi}-iM\omega{\hat r}r\right) \ \phi (r). \hfill \\ 
\end{gathered} 
\end{align}
From these two equations, we obtain 
\begin{align}\label{31}
\left( {{E^2} - {M^2}} \right)\phi (r) ={ \left[{\vec\sigma} \cdot\left( \vec\pi  + iM\omega\hat rr \right)\right] \left[{\vec\sigma} \cdot\left( \vec\pi  - iM\omega\hat rr \right)\right]}\phi(r).
\end{align}

Next we utilize $\left(\vec\sigma\cdot\vec a\right)\left(\vec\sigma\cdot\vec b\right)=\vec a\cdot\vec b+i\vec\sigma\cdot\left(\vec a\times\vec b\right)$ with $\vec a=\vec\pi  + iM\omega\hat rr $ and $\vec b=\vec\pi  - iM\omega\hat rr $, and note that $\vec a\cdot\vec b=\vec\pi\cdot\vec\pi-M^2\omega^2r^2$ (with $\vec\pi\cdot\vec\pi\phi(r)$ in Eq. \eqref{pidotpi}) and $\vec a\times\vec b=iM\omega r\left(\hat r\times\vec\pi-\vec\pi\times\hat r\right)$. We find $\left(\vec\pi\times\vec r - \vec r\times\vec\pi\right)\phi=-2\hat z\left(\frac m\alpha+\xi\right)\phi$ and $\left(\vec\pi\cdot\vec r - \vec r\cdot\vec\pi\right)\phi=-2i\phi$, so that Eq. \eqref{31} leads to an equation similar to Eq. (\ref{21}),
\begin{equation}\label{32}
\left[\frac{{{d^2}}}{{d{r^2}}} + \frac{1}{r}\frac{d}{{dr}} + {\kappa_\omega^2} - \frac{{{\eta ^2}}}{{{r^2}}}-M^2\omega^2r^2\right]\phi \left( r \right) = 0,
\end{equation}
with $\eta$ as in Eq. (\ref{22}) and
\begin{equation}\label{SimpleQ2}
\kappa_\omega^2=E^2-M^2+2M\omega\left[1+s\left(\frac m\alpha+\xi\right)\right].
\end{equation}
As $\omega$ approaches zero, Eq. (\ref{32}) reduces to Eq. (\ref{21}) with the only extra terms $-M^2\omega^2r^2$, and with Eq. (\ref{SimpleQ2}) which reduces to Eq. (\ref{22}).  With the change of variable $M\omega{r^2} = \rho$, then we rewrite the radial equation \eqref{32} in the form  
\begin{align}\label{33}
\rho\frac{{{d^2\phi (\rho) }}}{{d{\rho^2}}} +\frac{d\phi (\rho) }{{d\rho}} - \left( \frac{\eta^2}{4\rho}+\frac{\rho}{4}-\frac{\kappa_\omega^2}{4M\omega}\right)\phi (\rho) = 0.
\end{align}    
Let us write the solution of Eq. \eqref{33} in terms of the new function $F(\rho)$ defined as 
\begin{align}\label{35}
\phi (\rho ) = {\rho ^{\frac{{\left| {\eta} \right|}}{2}}}{e^{ - \frac{\rho }{2}}}F(\rho ){,}
\end{align}
so that Eq. \eqref{33} can be rewritten as
\begin{align}\label{36}
\rho F''(\rho ) + (1 + \left| {\eta} \right| - \rho )F'(\rho ) - \left(\frac{{1 + \left| {\eta} \right|}}{2} - \frac{{{\kappa_\omega^2}}}{{4M\omega }}\right)F(\rho ) = 0.
\end{align}  

The general solution for Eq. \eqref{35} is the confluent hypergeometric function  
\begin{align}\label{37}
\phi (\rho ) = {\rho ^{\frac{{\left| {\eta} \right|}}{2}}}{e^{ - \frac{\rho }{2}}}F\left(\left(\frac{{1 + \left| {\eta} \right|}}{2} - \frac{{{\kappa_\omega^2}}}{{4M\omega }}\right),1 +\eta, \rho \right).
\end{align}  
Even though Eq. (\ref{32}) clearly reduces to Eq. (\ref{21}) when $\omega\rightarrow 0$, as noted above, the corresponding reduction from the solution in Eq. (\ref{37}) (when expressed in terms of $r$) to Eq. (\ref{27}) is more subtle and we will not discuss it in detail. This is done by rescaling the solution of Eq. (\ref{32}) by a function of $M\omega$ (the factor of $r^2$ in the extra term $-M^2\omega^2r^2$ of Eq. (\ref{32})) so that it is still a solution and such that the limit $\omega\rightarrow 0$ exists and is non-zero. Hypergeometric functions allow for Taylor expansion and from this one can determine which expansion coefficients vanish or diverge in the limit. For divergent terms, one rescales by an appropriate power in $M\omega$ so that the limit exists and leads to the Bessel function of Eq. (\ref{27}). 

Because of the divergent behaviour of the function $ F(\rho) $ for large values of $\rho$, bound states solutions can only be obtained by imposing that this function becomes a polynomial of some degree $n$. Then the radial solution presents an acceptable behaviour at infinity. This condition is obtained by setting
\begin{align}\label{38}
\frac{{1 + \left| {\eta} \right|}}{2} - \frac{{{\kappa_\omega^2}}}{{4M\omega }} =  - n,\qquad n = 0,1,2,3, \dots
\end{align}
By substituting Eqs. (\ref{xi}), (\ref{22}) and (\ref{SimpleQ2}) into Eq. (\ref{38}), we obtain 
\begin{align}\label{new46}
E^2 - M^2=4M\omega \left[ n+\frac 1 2  \left| \frac { m }{ \alpha  } +\frac { 1 }{ 2\alpha  } \left( 1-\alpha  \right) s+\frac { aE }{ \alpha  }  \right| -\frac { s }{ 2 } \left( \frac { m }{ \alpha  } +\frac { 1 }{ 2\alpha  } \left( 1-\alpha  \right) s+ \frac { aE }{ \alpha  }  \right)  \right].
\end{align}
If the argument of the absolute value is positive, $\frac { m }{ \alpha  } +\frac { 1 }{ 2\alpha  } \left( 1-\alpha  \right) s+\frac { aE }{ \alpha  }>0$, then we find 
\begin{align}
E_{n,m}=  - \frac a{\alpha }M\omega(s - 1)  \pm \frac 1\alpha\sqrt {M\left[ {{a^2}M{{(s - 1)}^2}{\omega ^2} - \alpha\omega (2m + s)(s - 1) + {\alpha ^2}(M + 4n\omega  + (s - 1)s\omega )} \right]}.\label{new47}
\end{align}
Note that for $s=1$, the eigenvalues in Eq. \eqref{new47},
\begin{align}
 E_{n}=\pm\sqrt{M(M+4\omega n)},\label{newafter47}
\end{align}
are independent of $m$,  $a$, and $ \alpha$.  Fig. 1 displays $E_n$ from Eq. (\ref{newafter47}) as a function of $\omega$ for $M=1$ and various values of $n$. As per Eq. (\ref{newafter47}), Fig. 1 shows how $\left|E_n\right|$ increases when $n$ increases, for any $\omega$. Eq. (\ref{newafter47}) and Fig. 1 show that for small values of $\omega$, $E_n\rightarrow\pm M$ for any $n$. Eq. \eqref{new47} shows that $E_{n,m}$ depends on $n$ and the spectrum of energy is discrete.  

For $s=-1$, the expression for $E_{n,m}$ contains all the terms of Eq. (\ref{new47}) and depends on all its parameters, including $m$, $a$ and $\alpha$. The positive values of the energy spectrum with $s=-1$ is displayed as a function of $\alpha$ (with $M=1$, $m=1/2$, $a=0.1$ and $\omega=1$, for $n=1$ and $2$, in Fig. 2, and as a function of $\omega$ (with $M=1$, $m=1/2$, $a=0.1$ and $\alpha=0.2$) in Fig. 3. In both figures, we see that $\left|E\right|$ increases with $n$ for any value of $\alpha$ and $\omega$. For $a=0$, the energy spectrum of Eq. (\ref{new46}) is given by 
\begin{align}\label{42}
{ E }_{ n,m }=\pm \sqrt { { M }^{ 2 }+4M\omega \left[ n+\frac { 1 }{ 2 } \left| \frac { m }{ \alpha  } +\frac { 1 }{ 2\alpha  } \left( 1-\alpha  \right) s \right| -\frac { s }{ 2 } \left( \frac { m }{ \alpha  } +\frac { 1 }{ 2\alpha  } \left( 1-\alpha  \right) s \right)  \right]  },
\end{align} 
which is analogous to Eq. (28) in Ref. \cite{DO3} (we have considered the two-dimensional oscillator so that their $k=0$). 

When the argument of the absolute value is negative, $\frac { m }{ \alpha  } +\frac { 1 }{ 2\alpha  } \left( 1-\alpha  \right) s+\frac { aE }{ \alpha  }<0$, we obtain, rather than Eq. (\ref{new47}),
\begin{align}\label{49}
E_{n,m} =  - \frac a\alpha M\omega(s + 1)  \pm \frac 1\alpha \sqrt {{M\left[ {{a^2}M{{(s + 1)}^2}{\omega ^2} - \alpha\omega (2m + s)(s + 1) + {\alpha ^2}\left( {M+4n\omega  + (s + 1)s\omega } \right)} \right]}}.
\end{align}
Then, for $s=-1$, the eigenvalues in Eq. \eqref{49} are independent of $m$,  $a$, and $ \alpha$, as in Eq. (\ref{newafter47}). 
However, when $s=1$, the energy depends explicitly on all the parameters in Eq. (\ref{new47}), including $m$, $a$ and $\alpha$. Whether $s=-1$ or $+1$, $E$ depends on $n$ and the spectrum of energy is discrete.

\begin{figure}[H]\label{fig1}
 	\begin{center}
 		\includegraphics[scale=1]{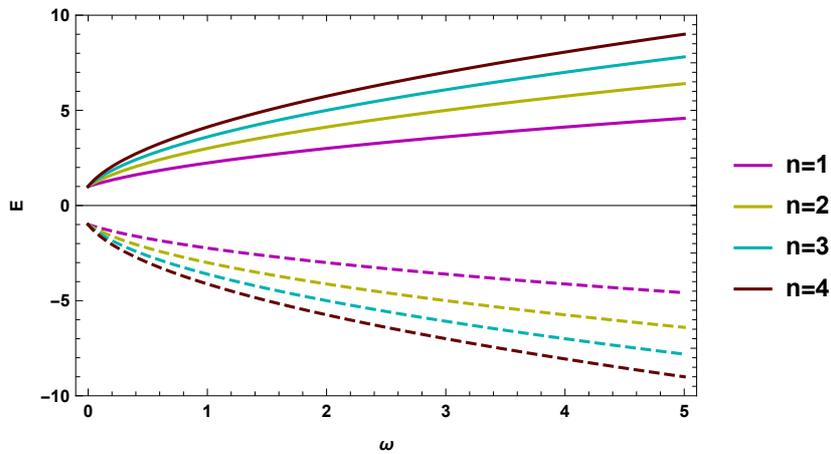}
 		\caption{Energy spectrum with $s=1$ as a function of $\omega$. The results are independent of $\alpha$ and $m$, with $M=1$ and different values of $n$.}
 	\end{center}
 \end{figure}

 \begin{figure}[H]\label{fig2}
 	\begin{center}
 		\includegraphics[scale=1]{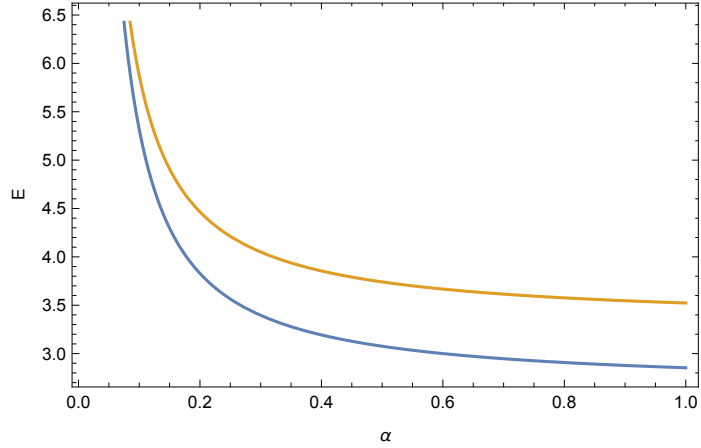}
 		\caption{Energy spectrum with $s=-1$ as a function of $\alpha$ with $M=1$, $m = 1/2$, $a=0.1$ and $\omega =1$ for $n=1$ (blue) and $n=2$ (dark yellow).}
 	\end{center}
 \end{figure}

 	\begin{figure}[H]\label{fig3}
 	\begin{center}
 		\includegraphics[scale=0.5]{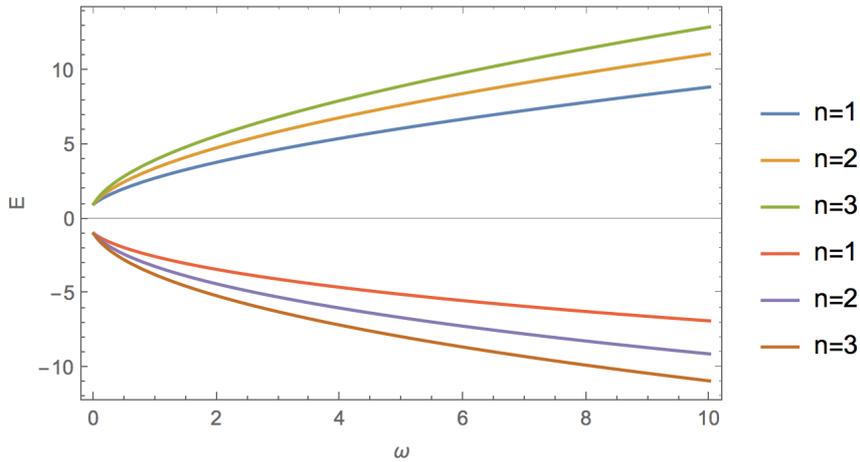}
 		\caption{Energy spectrum with $s=-1$ as a function of $\omega$, with $M=1$, $m=1/2$, $a=0.1$ and $\alpha=0.2$ for different values of $n$.
 		}
 	\end{center}
 \end{figure}


\section{Conclusion}

We have studied the relativistic quantum dynamics of a Dirac oscillator field subject to a linear interaction for spin-half particles in a cosmic string spacetime. The Dirac oscillator is a kind of tensor coupling with a linear potential which leads to the harmonic oscillator problem in the weak-coupling limit. This can be interpreted as the interaction of the anomalous magnetic moment with a linear electric field. From the corresponding Dirac oscillator equation, we analyzed the influence of the topological defect on the equation of motion, the energy spectrum and the wave-function.  Firstly, we solved the Dirac equation without the  oscillating term. We  obtained the Bessel function as a solution of the resulting equation. 

Next we turned to the Dirac oscillator, which is described by introducing a non-minimal coupling within the previous equations. The new term has direct implications on  the energy spectrum and wave-functions of the oscillator. The wave-function, which is a Bessel function without the Dirac oscillator, becomes, with the Dirac oscillator, related to the confluent hypergeometric function in Eq. (\ref{37}), and depends on the parameters $\eta$ (from Eq. (\ref{22})) and $\kappa_\omega$ (in Eq. (\ref{SimpleQ2})), thus depending on the parameters $m$, $\alpha$, $s$, $M$, $\omega$, and the energy $E$. In turn, we found that the energy eigenvalues $E$ depend on $m$, $\alpha$, $s$, $M$, $\omega$. Figs. 1 and 3 show that $\left|E\right|$ increases with $\omega$ for specific values of other parameters, and that $\left|E\right|$ increases with $n$. Fig. 2 shows that, for specific values of other parameters, $\left|E\right|$ decreases with $\alpha$ and increases with $n$.  The angular parameter $\alpha$ is related  to the linear mass density of the cosmic string. In the  limit $ \alpha=1$, that is, in the absence of topological defect,  we recover known results for the flat  spacetime. Here we have showed that the geometric and topological   properties  of these  spacetimes lead  to shifts in the energy spectrum and the wave-function that is comparable with the flat Minkowski spacetime.

\section*{Acknowledgement}

The authors thank Thomas Creutzig for helpful comments and the anonymous referee for helpful comments and suggestions on the manuscript. M. de Montigny acknowledges the Natural Sciences and Engineering Research Council (NSERC) of Canada for partial financial support (grant number RGPIN-2016-04309).


\begin{thebibliography}{99}


\bibitem{1} D. Ito, K. Mori, E. Carriere, {Nuov. Cim. A 51 (1967) 1119-1121. }
\bibitem{2} P.A. Cook, Lett. {Nuov. Cim. }1 (1971) 419-426.
\bibitem{3} H. Ui, G. Takeda, {Prog. Theor. Phys. 72 (1984) 266-284.}
\bibitem{4} M. Moshinsky, A. Szczepaniak, J. Phys. A{: Math. Gen. 22 (1989)  L817-L819.}
\bibitem{5} {R. P. Mart\'\i nez-y-Romero, A. L. Salas-Brito, J. Math. Phys, 33  (1992) 1831-1836.}
\bibitem{6} M. Moreno and A. Zentella, J. Phys. A : Math. Gen, 22 {(1989) L821-L825.} 
\bibitem{7} {D. Dutta, O. Panella, P. Roy, Ann.Phys. 331 (2013) 120-126.}
\bibitem{8} A. Bermudez, M.A. Martin-Delgado, A. Luis, Phys. Rev. A 77 {(2008) 063815(13pp).}
\bibitem{9} A. Bermudez, M.A. Martin-Delgado, E. Solano, Phys. Rev. A 76 {(2007)  041801(R,4pp).}
\bibitem{11} {J. Ben\'\i tez, R. P. Mart\'\i nez y Romero, H. N. N\~ unez-Y\'epez, A. L. Salas-Brito, Phys. Rev. Lett. 64 (1990) 1643-1645.}
\bibitem{12} {O. Casta\~nos, A. Frank, R. L\'opez, L. F. Urrutia, Phys. Rev. D 43 (1991) 544-547.}
\bibitem{13} J. Grineviciute, D. Halderson, Phys. Rev. C 80 {(2009) 044607(8pp).}
\bibitem{14}  {M. Moshinsky, Y.F. Smirnov,} The Harmonic Oscillator in Modern Physics, Harwood Academic Publishers, Amsterdam, 1996.
\bibitem{15}  {M. Moshinsky,} G. Loyola, Found. Phys. 23 (1993) 197-210.
\bibitem{16} {R. de Lima Rodrigues, Phys. Lett. A 372 (2008) 2587-2591. }
\bibitem{17} J. P. Crawford, J. Math. Phys. {34  (1993) 4428-4435.}
\bibitem{18} F. Vega, J. Math. Phys. {55  (2014) 032105(8pp). } 
\bibitem{19} S. Cai, T. Jing, G. Guo, R. Zhang, Int. J. Theor. Phys. 49 {(2010) 1699-1705}. 
\bibitem{20} M. H. Pacheco, R. R. Landim, C. A. S. Almeida, { Phys. Lett. A 311 (2003) 93-96}.
\bibitem{21} C. Quesne, M. Moshinsky, J. Phys. A{: Math. Gen.  23} (1990) 2263-2272.
\bibitem{22} {J. Beckers, N. Debergh}, Phys. Rev. D 42 (1990) 1255-1259.
\bibitem{23} R. Rekioua, T. Boudjedaa, Eur. Phys. J. C 49 (2007) 1091-1098.
\bibitem{25} B. P. Mandal, S. Verma, Phys. Lett. A 374 (2010) 1021-1023.
\bibitem{26} N. Ferkous, A. Bounames, Phys. Lett. A 325 (2004) 21-29.
\bibitem{27} E. Sadurn\'\i, AIP Conf. Proc. 1334 (2011) 249-290.
\bibitem{28} H Hassanabadi, M Hosseinpour{, M de Montigny,  Int. J. Mod. Phys. A 31 (2016) 1650191(19pp). }
\bibitem{29} H Hassanabadi, M Hosseinpour{, M de Montigny,  Eur. Phys. J. Plus 132 (2017) 541(12pp). }
\bibitem{DO1} K. Bakke, Gen. Relat. Grav. 45 (2013) 1847-1859.
\bibitem{DO2} K. Bakke, Eur. Phys. J. Plus 127 (2012) 82(8pp).
\bibitem{DO3} J. Carvalho, C. Furtado, F. Moraes, Phys. Rev. A 84 (2011) 032109(6pp).
\bibitem{DO4} P. Strange, L. H. Ryder, Phys. Lett. A 380 (2016) 3465-3468.
\bibitem{DO5} R.L.L. Vit\'oria, K. Bakke, Eur. Phys. J. C 78 (2018) 175(6pp).
\bibitem{DO6} K. Bakke, C. Furtado, Ann. Phys. 336 (2013) 489-504.
\bibitem{DO7} K. Bakke, H. Mota, Eur. Phys. J. Plus 133 (2018) 409(9pp).
\bibitem{SCA1} A. Boumali, N. Messai, Can. J. Phys. 92 (2014) 1460-1463.
\bibitem{SCA2} R.L.L. Vit\'oria, K. Bakke, Eur. Phys. J. Plus 131 (2016) 36(8pp).
\bibitem{SCA3} L.C.N. Santos, C.C. Barros Jr, Eur. Phys. J. C 78 (2018) 13(8pp). 
\bibitem{SCA4}  M. Hosseinpour, H. Hassanabadi, Eur. Phys. J. Plus 130 (2015) 236(10pp).  
\bibitem{SCA5}  L.B. Castro, Eur. Phys. J. C 76 (2016) 61(11pp).   
\bibitem{RO1}  J. Carvalho, A. M. de M. Carvalho, E. Cavalcante, C. Furtado, Eur. Phys. J. C 76 (2016) 365 (9pp).
\bibitem{RO2} K. Bakke, C. Furtado, Phys. Lett. A 376 (2012) 1269-1273.
\bibitem{RO3} L.F. Deng, C. Y. Long, Z. W. Long, T. Xu, Adv. High En. Phys. (2018) 2741694 (10pp).
\bibitem{RO4} D. Nath, P. Roy, Ann. Phys. 351 (2014) 13-21.
\bibitem{RO5} R.R.S.Oliveira, R.V.Maluf, C.A.S.Almeida, Ann. Phys. 400 (2019) 1-8.
\bibitem{RO6}  B.P. Mandal, S. K. Rai, Phys. Lett. A 376 (2012) 2467-2470.
\bibitem{RO7} R.L.L. Vit\'oria, K. Bakke, Int. J. Mod. Phys. D 27 (2018) 1850005.
\bibitem{RO8} R.L.L. Vit\'oria, H. Belich, K. Bakke, Eur. Phys. J. Plus 132 (2017) 25 (7pp).
\bibitem{RO9} A. Dhar, K. Wagh, Europhys. Lett. 79 (2007) 60003 (5pp).
\bibitem{Cunha2016} M. S. Cunha, C. R. Muniz, H. R. Christiansen, V. B. Bezerra, Eur. Phys. J. C 76 (2016) 512 (7pp).
\bibitem{30} P. O. Mazur, Phys. Rev. Lett. 57 (1986) 929-932.  
\bibitem{31} C.R. Muniz, V.B. Bezerra, M.S. Cunha, Ann. Phys. 350 (2014) 105-111.
\bibitem{32} V.B. Bezerra, J. Math. Phys. 38 (1997) 2553-2564.
\bibitem{33} J.D. Bekenstein, Phys. Rev. D 45 (1992) 2794-2801.
\bibitem{34} N. D. Birrel and P. C. W. Davies, Quantum Fields in Curved Space, Cambridge University Press, Cambridge, England, {1984}.
\bibitem{35} M. Nakahara, Geometry, Topology and Physics {2nd Ed., Institute of Physics Publishing, Bristol, 2003}.
\bibitem{36} H. Hassanabadi, M. Hosseinpour, Eur. Phys. J. C {76 (2016) 553(7pp).}
\bibitem{37} M Hosseinpour{, F. M. Andrade, E.O. Silva, H. Hassanabadi, Eur. Phys. J. C 77 (2017) 270(6pp); erratum ibid. 373.} 
\bibitem{CaiPapini} Y.Q. Cai, G. Papini, Class. Quant. Grav. 7, 269 (1990). 
\bibitem{Vilenkin1985} A. Vilenkin, Phys. Rep. 121, 263 (1985).
\bibitem{andrade} F.M. Andrade, E. O. Silva, Eur. Phys. J. C 74 (2014) 3187(8pp).


\end{thebibliography}
\end{document}